\documentclass[ reprint, amsmath, amssymb, aps, prmaterials, longbibliography]{revtex4-2}
\usepackage{booktabs}
\usepackage{longtable}
\usepackage{multirow}

\usepackage{graphicx}
\usepackage{dcolumn}
\usepackage{bm}
\usepackage[hidelinks]{hyperref}

\begin{document}
	

\title{Wyckoff-Resolved Oxidation-State Atlas and Anion-Conditioned Priors for Materials Discovery} 

\author{Boris Kiefer} 
\email{bkiefer@nmsu.edu} 
\affiliation{Department of Physics, New Mexico State University, Las Cruces, New Mexico 88003, USA} 

\date{\today} 

\begin{abstract}
	We introduce a Wyckoff-resolved oxidation-state atlas and assignment utility for probability-ranked, charge-neutral assignment from compositions or Wyckoff grammars. The atlas is constructed from a May 2026 snapshot of \(154{,}879\)
	Materials Project-derived structures by staged exact-neutral enumeration over common and known nonzero oxidation states. The learned prior assigns \(106{,}053\) materials in composition mode and \(114{,}403\) in Wyckoff mode, compared with a broad MP all-integer \texttt{possible\_species} count of \(108{,}642\). A matched MP baseline requiring at least one charge-neutral assignment with exactly one nonzero integer oxidation state per element contains \(89{,}374\) materials; relative to this baseline, composition and Wyckoff modes increase coverage by \(18.7\%\) and \(28.0\%\). Of the \(14{,}665\) materials recovered only in Wyckoff mode, \(99.98\%\) exhibit distinct formal oxidation states for the same element on different site tokens. The CSV/Python workflow provides a reproducible prior for structure decoration, generative crystal models, and symbolic Wyckoff-grammar workflows.
\end{abstract}

\keywords{oxidation states, Wyckoff positions, crystal chemistry, materials informatics, charge neutrality, generative materials models}



\maketitle

\section{Introduction}

Charge neutrality is one of the simplest and most widely used constraints in inorganic crystal chemistry. It underlies the interpretation of ionic solids, polar functional oxides, solid electrolytes, and semiconducting compounds, even when the electronic structure is far from a purely ionic limit. In high-throughput materials databases such as the Materials Project (MP), this constraint is routinely combined with crystallographic and electronic-structure information to organize large collections of computed inorganic materials \cite{Jain2013MaterialsProject}. Modern materials-informatics workflows increasingly rely on explicit structural representations, graph neural networks, and generative models to propose or evaluate hypothetical crystals \cite{Ong2013Pymatgen,Xie2018CGCNN,Reiser2022GNNReview,Merchant2023GNoME,DiffCSP2023}. Wyckoff sequences provide a finite, enumerable, coordinate-free encoding of crystal symmetry across all 230 space-group types \cite{hornfeck2022combinatorics}, and have been used for machine-learning property prediction \cite{goodall2022rapid} and symmetry-aware generative crystal models \cite{WyCryst2023,CrystalFormer2024,WyckoffDiff2025}. For such workflows, chemically meaningful charge-balance constraints can serve as fast filters, priors, and diagnostics.

Formal oxidation states provide a compact way to impose charge neutrality. Bond-valence methods offer a chemically interpretable route for relating local coordination and bond lengths to formal valence \cite{Brown1985BondValence,Brese1991BondValence}, while charge partitioning and spectroscopic approaches can provide complementary oxidation-state information from experiment or electronic-structure analysis \cite{Bader1990AIM,Henkelman2006Bader,Manz2010DDEC,Yano2009XAS}. Materials-informatics software such as \texttt{pymatgen} provides practical tools for oxidation-state decoration and structure analysis \cite{Ong2013Pymatgen}, and large databases including the Materials Project and AFLOW expose computed structures and metadata that enable large-scale oxidation-state analysis \cite{Jain2013MaterialsProject,Curtarolo2012AFLOW}.

Composition-based oxidation-state assignment is powerful, but it becomes ambiguous when the same element can adopt more than one formal oxidation state in a structure. For example, \(\mathrm{Fe_2O_3}\) is naturally described by \(\mathrm{Fe}^{+3}\) and \(\mathrm{O}^{-2}\), whereas \(\mathrm{Fe_3O_4}\) has the composition-averaged Fe oxidation state \(+8/3\) even though the conventional chemical picture involves both \(\mathrm{Fe}^{+2}\) and \(\mathrm{Fe}^{+3}\) on distinct Wyckoff sites \cite{verwey1941electronic,wright2001long}. This ambiguity is common in transition-metal compounds and related functional materials, including battery materials such as \(\mathrm{LiFePO_4}\), where the \(\mathrm{Fe}^{+2}/\mathrm{Fe}^{+3}\) redox couple and small-polaron transport are central to electrochemical operation \cite{Padhi1997LiFePO4,Maxisch2006LiFePO4Polaron,Manthiram2009PhosphoOlivine}; catalytic oxides with variable redox-active sites \cite{Wachs2012MixedOxideCatalysis,Sahoo2024TMOReview}; correlated electron systems in which charge order, mixed valence, and strong correlations control emergent properties \cite{Dagotto2001CMR,Tokura2000OrbitalPhysics}; multiferroics where site-specific valence configurations can contribute to magnetoelectric coupling \cite{Eerenstein2006Multiferroics,VanDenBrink2008ChargeOrderMultiferroics}; and transparent conducting or spinel oxides where cation choice, defect chemistry, and mixed-metal chemistry control transport and optical response \cite{Walsh2011TCO,Amini2014SpinelTCO,Zhang2016PTypeTCO}.

Oxidation-state ambiguity is not restricted to cations. Nitrogen can occur as \(\mathrm{N}^{-3}\) in nitrides, in lower negative states in azides and related species, and in positive formal oxidation states in oxo-nitrogen environments; sulfur similarly spans sulfide, sulfite, sulfate, and related chemistries \cite{GreenwoodEarnshaw1997,Sun2019TernaryNitrides,Kageyama2018MixedAnion}, and many combinations of formal charges can satisfy the same global charge-neutrality condition. This ambiguity matters for materials discovery because generative and screening workflows often produce candidate structures, decorations, or Wyckoff grammars before detailed bond lengths, local relaxations, or full electronic-structure calculations are available. A lightweight oxidation-state prior can reject impossible assignments, rank chemically plausible decorations, flag unusual oxidation states, and identify site-resolved formal oxidation-state patterns that composition-only rules cannot represent. Such a prior should be transparent, reproducible, and easy to combine with graph-based or grammar-based materials representations.

Here we construct a global, Wyckoff-resolved oxidation-state atlas from a large corpus of inorganic crystal structures cataloged in the MP database \cite{Jain2013MaterialsProject}. Instead of returning a single oxidation assignment per material, the atlas tabulates element-normalized empirical probabilities for integer oxidation states. 
The candidate oxidation-state sets were initialized from the tabulation
of common and known oxidation states in Greenwood and Earnshaw's
\emph{Chemistry of the Elements} \cite{GreenwoodEarnshaw1997}. These
states were organized into a staged level-0 and level-1 exact-neutral
enumeration strategy: common oxidation states are used first, and
additional known oxidation states are introduced only for structures
that fail the common-state stage.
This produces a transparent, user-customizable prior that can be applied to both composition-derived analysis, similar in spirit to existing composition- and software-based oxidation-state tools \cite{Ong2013Pymatgen,Fu2023BERTOS,Thway2024OSP}, and the Wyckoff-grammar structure representations introduced above.

The main contribution is the use of this atlas as an assignment engine. We first compare atlas-prior assignments with the broad MP all-integer \texttt{possible\_species} classification, which provides a database-scale coverage benchmark but does not by itself require a charge-neutral assignment with one nonzero integer oxidation state per element. Composition mode reaches \(106{,}053\) materials, close to the broad MP count of \(108{,}642\); Wyckoff mode reaches \(114{,}403\). For the mechanistic uplift analysis, we therefore construct a stricter matched MP composition baseline containing \(89{,}374\) materials that satisfy the same one-state-per-element, nonzero, integer, charge-neutral criterion used for the atlas composition comparison. Relative to this matched baseline, the composition-level uplift isolates oxidation-state-library and enumeration effects, while the additional Wyckoff uplift isolates the effect of assigning crystallographically distinct site tokens independently. Nearly all materials assigned by Wyckoff mode but not by composition mode contain at least one element assigned different formal oxidation states on distinct Wyckoff/site tokens.

\section{METHODS}
\label{sec:methods}

\subsection{Materials Project input structures and Wyckoff grammar}

We constructed the input set from a snapshot of MP entries with DFT-relaxed structures and associated metadata retrieved in May 2026, giving \(N=154{,}879\) materials. Because the Materials Project database is continuously updated, this count refers to the archived input snapshot used throughout the present analysis rather than to
the current live database. AFLOW tools were used for symmetry analysis and generation of CIF files for Wyckoff-grammar extraction \cite{hicks2018aflow}. Symmetry was assigned by comparing AFLOW-determined and MP-reported space groups. A first pass used ``tight'' symmetry constraints; structures whose AFLOW space group was lower than the MP value were reprocessed with ``loose'' symmetrization. After these two stages, \(154{,}280/154{,}879\) structures, or approximately \(99.4\%\), were converted successfully from POSCAR/CONTCAR format to Wyckoff-aware CIF representations. For the remaining \(599\) materials, the AFLOW-determined symmetry was retained.

The resulting materials information was serialized into a parsable CSV row format containing the MP identifier, reduced formula, band gap, energy above hull, space-group number and symbol, number of atomic sites, and compact Wyckoff-site grammar. The \texttt{sites} field encodes element, Wyckoff multiplicity, Wyckoff letter, and site label for each crystallographic site token, following the space-group and Wyckoff-position conventions of the \emph{International Tables for Crystallography A} \cite{InternationalTablesA}. For example, \texttt{O:8:f:O3} denotes an oxygen site on a Wyckoff position of multiplicity \(8\) and letter \(f\). This representation allows the same element to occupy multiple crystallographically distinct site tokens and is therefore richer than composition-only oxidation-state assignment.

\subsection{Oxidation-state levels}

Candidate oxidation states were defined from a user-editable element table containing atomic number, element symbol, common oxidation states, and additional known oxidation states. The table was initialized from Greenwood and Earnshaw's \emph{Chemistry of the Elements} \cite{GreenwoodEarnshaw1997}. The common set is denoted level-0 and the additional known set level-1. The complete oxidation-state table is provided in the Supplemental Material~\cite{SupplementalMaterial}. For element \(Z\), these sets are \(\Omega_0(Z)\) and \(\Omega_1(Z)\), and the level-union candidate set is
\begin{equation}
	\Omega_{\leq L}(Z) =
	\bigcup_{\ell=0}^{L} \Omega_{\ell}(Z).
	\label{eq:level_union}
\end{equation}

The atlas was built with a staged level-0/level-1 workflow. Level-0 contains common nonzero oxidation states, while level-1 adds less common but chemically known nonzero states. Zero-valent states were excluded because they can trivially satisfy charge neutrality in metallic or covalent systems and would obscure the oxidation-state statistics targeted here.

\subsection{Staged exact-neutral atlas construction}

For each material, we searched for integer oxidation-state assignments satisfying exact charge neutrality. In Wyckoff mode, each crystallographic site token is an independent assignment variable. If site token \(i\) has element \(Z_i\), multiplicity \(m_i\), and oxidation state \(x_i\), exact neutrality requires
\begin{equation}
	\sum_i m_i x_i = 0,
	\label{eq:charge_neutrality}
\end{equation}
with
\begin{equation}
	x_i \in \Omega_{\leq L}(Z_i).
	\label{eq:allowed_oxidation_state}
\end{equation}

Level 0 was evaluated first. Materials with at least one exact-neutral level-0 solution were recorded as level-0 successes; failures were passed to level 1, where the allowed set was expanded to \(\Omega_{\leq 1}(Z)\). This makes the level-0 and level-1 recovery sets disjoint and allows level 1 to be interpreted as the incremental recovery from adding chemically known but less common oxidation states.

Each stage used a fixed combination limit to prevent uncontrolled brute-force enumeration. Materials whose candidate product space exceeded this limit were recorded as combination-limit exceeded; materials with no exact-neutral solution within the allowed candidate space were recorded separately. Single-element materials were counted but excluded from atlas construction because formal oxidation-state assignment is not meaningful for elemental phases in the same way as for multi-element compounds.

For each retained exact-neutral solution, oxidation-state counts were accumulated by element and oxidation state. We tracked material-level occurrence, Wyckoff-site occurrence, and Wyckoff-position occurrence. Unless otherwise stated, reported atlas probabilities use Wyckoff-site mass, which preserves crystallographic multiplicity while avoiding the stronger size dependence of Wyckoff-position counts.

\subsection{Atlas probabilities}

Level-0 and level-1 solution tallies were accumulated into a union atlas. For each element \(Z\) and oxidation state \(q\), the atlas contains a nonnegative mass \(M(Z,q)\). The element-normalized probability is
\begin{equation}
	P(q \mid Z) =
	\frac{M(Z,q)}
	{\sum_{q'} M(Z,q')}.
	\label{eq:atlas_probability}
\end{equation}

These probabilities define the oxidation-state prior used by the assignment utility. Normalization is performed independently for each element, so the atlas encodes relative oxidation-state probabilities for a given element rather than an overall element-abundance prior.

We constructed a global ALL atlas and O-, N-, and S-conditioned atlases. The global atlas uses the full multi-element structure set. The conditioned atlases filter the input to materials containing at least one O, N, or S atom, respectively, and then repeat the same staged tallying and normalization procedure. These conditioned atlases assess chemical-family dependence and are reported in the Supplemental Material~\cite{SupplementalMaterial}.

\subsection{Atlas-prior assignment utility}
\label{sec:methods_assignment}

The union atlas was used as a prior for assigning oxidation states beyond the exact-enumeration construction subset. Candidate states were filtered by a minimum atlas probability,
\begin{equation}
	P(q \mid Z) \geq P_{\min}.
	\label{eq:pmin_filter}
\end{equation}
Unless otherwise stated, \(P_{\min}=0.01\), corresponding to a \(1\%\) element-normalized threshold (see also main text Table~\ref{tab:pmin_ox}). This removes low-probability tails while retaining most observed element--oxidation-state support; retained probabilities were not renormalized.

Two assignment modes were implemented. In composition mode, the formula was parsed with \texttt{pymatgen}, and each element was assigned one oxidation state. Charge neutrality was evaluated as
\begin{equation}
	\sum_Z n_Z x_Z = 0,
	\label{eq:composition_charge_neutrality}
\end{equation}
where \(n_Z\) is the stoichiometric coefficient of element \(Z\). In Wyckoff mode, each crystallographic site token was assigned independently, and charge neutrality was evaluated using Eq.~\eqref{eq:charge_neutrality}; thus, distinct sites of the same element can carry different oxidation states.

For each element or Wyckoff/site variable, candidate oxidation states were sorted in descending order of \(P(q \mid Z)\). Candidate tuples were then evaluated in the direct Cartesian-product order of these probability-ranked per-variable lists. This traversal favors high-probability states locally but is not equivalent to globally sorting all candidate tuples by their total assignment prior. It also does not seed or otherwise prioritize composition-equivalent site-uniform assignments in Wyckoff mode. For each tested tuple, we recorded the total charge, assignment string, sum of log prior probabilities,
\begin{equation}
	\log P_{\mathrm{assign}} =
	\sum_i \log P(x_i \mid Z_i),
	\label{eq:assignment_log_prior}
\end{equation}
and diagnostics such as minimum and mean site probability. Assignment caps of \(N_{\mathrm{assign}}=10^{4}\) and \(N_{\mathrm{assign}}=10^{5}\) were used to test search-budget dependence (see also Sec.~\ref{sec:benchmarking}).

Two search modes were used. Fast-first mode stops when the first exact-neutral assignment is found or the cap is reached. Exhaustive-ranked mode tests all assignments up to the same cap, collects all exact-neutral assignments encountered, and reports the highest-prior neutral assignment. For a fixed candidate order and cap, both modes identify the same material-level success set, but they can report different assignments when multiple neutral solutions occur within the cap.

\subsection{Implementation and reproducibility}
All analysis steps were implemented as plain-text Python workflows operating on CSV inputs and outputs. The workflow consists of generating the master Wyckoff grammar file, performing staged level-0 and level-1 exact-neutral enumeration, accumulating the union atlas, visualizing element-normalized oxidation-state probabilities, applying the atlas-prior assignment utility in composition and Wyckoff modes, comparing against MP \texttt{possible\_species}, and analyzing the mixed-Wyckoff mechanism of the assignment uplift. The scripts write run summaries, filtered atlas probability tables, atlas-support statistics, material-level assignment files, and set-comparison summaries to make the analysis auditable and reproducible.

In addition to the corpus-scale workflow, the repository provides a
standalone command-line utility for applying the trained atlas to
user-supplied materials. The utility accepts either individual
compositions or Wyckoff-resolved site grammars and returns
probability-ranked exact-neutral assignments together with assignment
weights and site-probability diagnostics. Representative calculations for \(\mathrm{MgO}\), \(\mathrm{SnO}_{2}\), and \(\mathrm{Fe}_{3}\mathrm{O}_{4}\), including unique, ambiguous, composition-unresolved, and Wyckoff-resolved outcomes, are provided in the Supplemental Material~\cite{SupplementalMaterial}.

ChatGPT (OpenAI; multiple model versions used during development) was used to assist with Python code development. The computational workflow, analysis design, and scientific interpretation were developed and directed by the author. All AI-assisted code was reviewed, tested, and verified by the author, and the resulting outputs were independently checked before submission.

\section{Results and discussion}

We applied the staged level-0/level-1 oxidation-state workflow described in  Sec.~\ref{sec:methods} to the MP-derived serialized Wyckoff grammar dataset. The resulting ALL atlas is used for the main-text analysis. The O-, N-, and S-conditioned atlases were generated with the same procedure and are reported in the Supplemental Material~\cite{SupplementalMaterial}. This organization separates the global, MP-comparable analysis from chemically conditioned anion-family diagnostics.

A central distinction is between \emph{atlas construction} and \emph{atlas application}. As described in Methods, atlas construction uses only materials for which the staged enumeration workflow produces exact charge-neutral solution ensembles. These ensembles define the empirical masses \(M(Z,q)\) and element-normalized prior \(P(q \mid Z)\). Atlas application then applies this learned prior back to the original MP-derived master list, requiring discovery of at least one exact-neutral assignment within a probability-filtered candidate space and fixed search budget. Thus, atlas application can recover more assigned materials than the ensemble-resolved subset used to construct the atlas. The distinction between the ensemble-resolved construction set and
the larger atlas-application success set demonstrates that a
conservative construction procedure can yield a transferable prior
for broader ranked assignment.

\subsection{Atlas construction and training}

The atlas was constructed using Wyckoff-site variables, adding a minimal crystallographic layer beyond composition-only oxidation-state assignment. Table~\ref{tab:ALL_stats} summarizes the staged construction accounting. Level 0 is applied to the full master list, with level 1 applied only to materials unresolved at level 0. The rows therefore report both material-level outcomes and the accumulated Wyckoff-site solution weight used to construct the probability atlas.
\begin{table*}[t]
	\caption{Separated level-0 and level-1 accounting for global atlas construction. Level-0 is applied to the full master list, while level-1 is applied only to materials not resolved at level-0. Material-level rows report construction outcomes; the final row reports the accumulated Wyckoff-site solution weight used to construct the atlas probabilities.}
	\label{tab:ALL_stats}
	\begin{ruledtabular}
	\begin{tabular}{lcc}
		property & level-0 & level-1 \\
		\hline
		Input materials & 154,858 & 92,823 \\
		Input origin & master list & level-0 failed \\
		Missing oxidation states & 3 & 3  \\
		Single-element materials & 827 & 827 \\
		$N_{\mathrm{assign}}$ exceeded & 16,577 & 40,012 \\
		Exact-neutral solution & 62,035 & 27,431 \\
		No exact-neutral solution & 73,120 & 23,025 \\
		Post-filtered neutral solutions & 2,290 & 1,519 \\
		\hline
		Accounting sum & 154,858 & 92,823 \\
		Unresolved after level & 92,823 & 65,392 \\
		\hline
		Wyckoff-site solution weight & 716,503 & 121,940 \\
	\end{tabular}
	\end{ruledtabular}
\end{table*}
The table separates chemical recovery from bookkeeping outcomes. A small number of entries contain elements without allowed oxidation states in the level table, and single-element entries are counted but excluded from atlas construction. The \(N_{\mathrm{assign}}\)-exceeded and no-solution rows identify different unresolved cases: the former exceed the enumeration cap, whereas the latter are searched within the allowed candidate space but do not yield an exact-neutral assignment. Among solved materials, level 0 accounts for \(69\%\) and level 1 for \(31\%\), confirming that common oxidation states dominate while additional known states provide substantial recovery. The final Wyckoff-site solution weight, approximately \(8.4\times10^{5}\), is the statistical support used to define the Wyckoff-resolved atlas probabilities. 

\begin{figure*}[p]
	\centering
	\includegraphics[width=\textwidth, height=0.9\textheight]{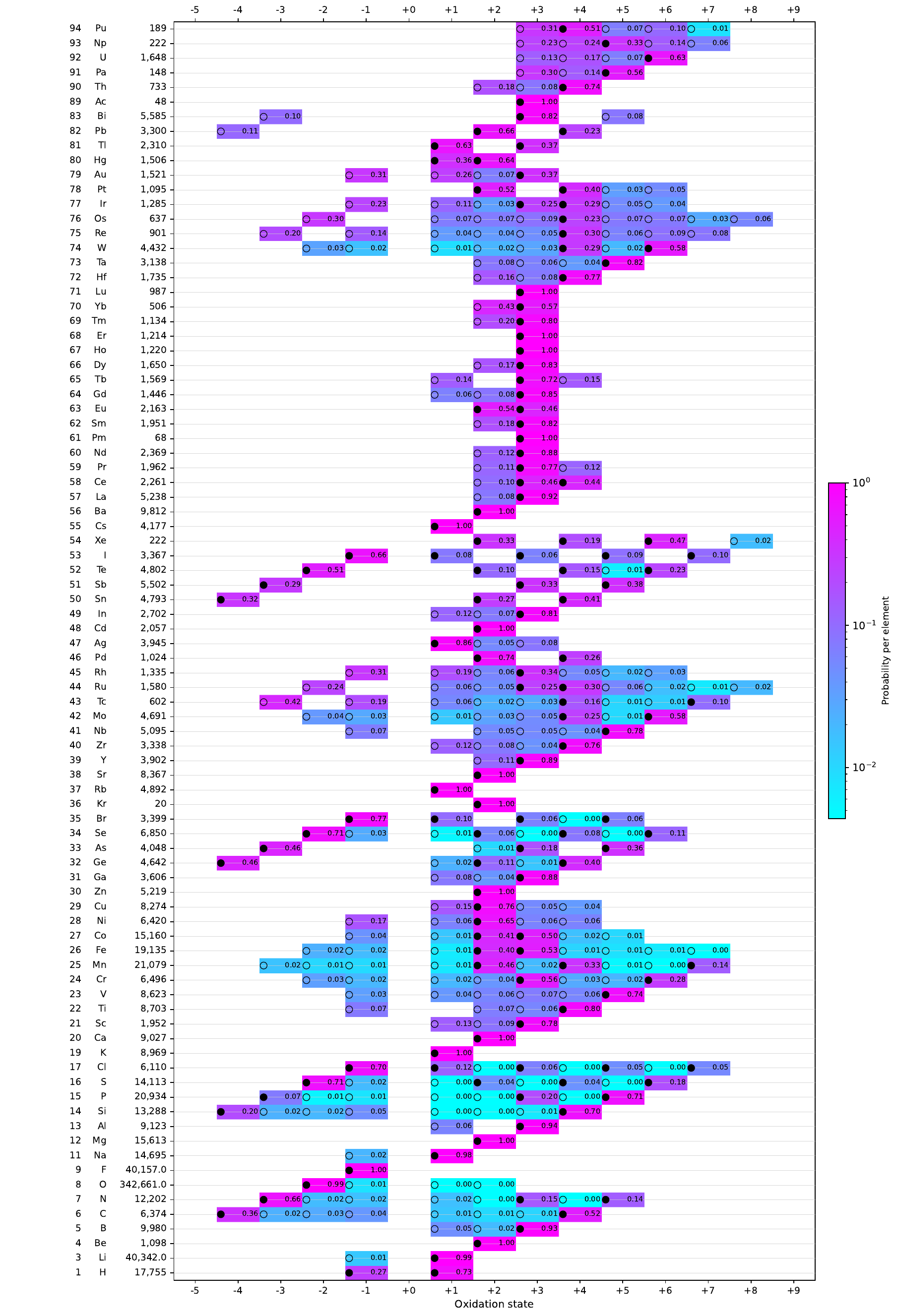}
	\caption{Global oxidation-state atlas. Each row corresponds to an element and each column to an integer oxidation state. Colors denote the element-normalized probability \(P(q \mid Z)\) obtained from the staged level-0/level-1 atlas construction. Filled markers indicate level-0 oxidation states and open markers indicate level-1 oxidation states. The left annotation reports the total Wyckoff-site occurrence for each element.}
	\label{fig:heatmap_all}
\end{figure*}

The global ALL-atlas heatmap (Fig.~\ref{fig:heatmap_all}) provides the oxidation-state prior used by the assignment utility. Strongly ionic species are concentrated in familiar oxidation states, such as \(\mathrm{O}^{-2}\), while transition metals retain broader support across multiple positive states. The level-1 tail is chemically useful for recovering unusual or site-resolved assignments, but must be filtered during ranked assignment to prevent low-probability states from dominating the combinatorial search.

\subsection{Property-resolved construction diagnostics}

We next tested whether the sequential level-0\(\rightarrow\)level-1 construction produces a property-trivial subset of the MP-derived corpus. For each bandgap and energy-above-hull bin, level-0 successes are retained and level-0 failures are replaced by their level-1 outcomes.

Tables~\ref{tab:bandgap_accounting} and~\ref{tab:ehull_accounting} show complementary behavior. Exact-neutral success increases from \(45.8\%\) for \(E_g \le 10^{-5}\,\mathrm{eV}\) to \(77.9\%\) for \(E_g > 2\,\mathrm{eV}\), while the no-solution fraction decreases from \(29.7\%\) to \(0.1\%\). This trend is consistent with the greater compatibility of discrete formal oxidation-state assignments with finite-gap compounds. In contrast, energy-above-hull stratification is weaker: success ranges from \(60.5\%\) for \(E_{\mathrm{hull}}\le 0.05\,\mathrm{eV}\) to \(52.9\%\) for \(E_{\mathrm{hull}}>0.25\,\mathrm{eV}\). Thus, the atlas is sensitive to chemically meaningful electronic character without acting simply as a near-stable materials filter.
\begin{table*}[t]
	\caption{
		Band-gap-resolved accounting of oxidation-state enumeration outcomes for the all-compound sequential level-0\(\rightarrow\)level-1 atlas at \(N_{\mathrm{assign}}=10^{4}\), using the assignment-cap definition in Sec.~\ref{sec:methods_assignment}.
		Level-0 is applied first, and level-1 is applied only to entries that fail to obtain an exact-neutral level-0 assignment; therefore, level-0 successes are retained and level-0 failures are replaced by level-1 outcomes.
		Percentages are normalized within each band-gap bin and sum to \(100\%\).
		The ``Other'' category groups minor bookkeeping outcomes.
	}
	\label{tab:bandgap_accounting}
	\begin{ruledtabular}
	\begin{tabular}{lrrrrr}
		\toprule
		Band-gap bin & $N$ & Success & No solution & Limit & Other \\
		\hline
		$E_g \le 10^{-5}$        & 72,636 & 45.8 & 29.7 & 22.0 & 2.5 \\
		$10^{-5} < E_g \le 1$    & 27,067 & 56.4 & 4.7  & 37.9 & 1.0 \\
		$1 < E_g \le 2$          & 18,489 & 66.9 & 0.6  & 31.9 & 0.6 \\
		$E_g > 2$                & 36,666 & 77.9 & 0.1  & 21.6 & 0.4 \\
		\botrule
	\end{tabular}
	\end{ruledtabular}
\end{table*}
\begin{table*}[t]
	\caption{
		Energy-above-hull-resolved accounting of oxidation-state enumeration outcomes for the all-compound sequential level-0\(\rightarrow\)level-1 atlas at \(N_{\mathrm{assign}}=10^{4}\), using the assignment-cap definition in Sec.~\ref{sec:methods_assignment}.
		Level-0 is applied first, and level-1 is applied only to entries that fail to obtain an exact-neutral level-0 assignment; therefore, level-0 successes are retained and level-0 failures are replaced by level-1 outcomes.
		Percentages are normalized within each \(E_{\mathrm{hull}}\) bin and sum to \(100\%\).
		The ``Other'' category groups minor bookkeeping outcomes.
	}
	\label{tab:ehull_accounting}
	\begin{ruledtabular}
	\begin{tabular}{lrrrrr}
		\toprule
		$E_{\mathrm{hull}}$ bin & $N$ & Success & No solution & Limit & Other \\
		\hline
		$E_{\mathrm{hull}} \le 0.05$        & 77,895 & 60.5 & 17.1 & 20.7 & 1.7 \\
		$0.05 < E_{\mathrm{hull}} \le 0.10$ & 25,741 & 57.7 & 8.5  & 32.9 & 0.9 \\
		$0.10 < E_{\mathrm{hull}} \le 0.25$ & 27,462 & 54.4 & 12.2 & 32.4 & 1.0 \\
		$E_{\mathrm{hull}} > 0.25$          & 23,760 & 52.9 & 17.4 & 27.4 & 2.2 \\
		\botrule
	\end{tabular}
	\end{ruledtabular}
\end{table*}

\subsection{Atlas benchmarking and matched MP comparison}
\label{sec:benchmarking}

We benchmarked atlas-prior assignments against two related MP
\texttt{possible\_species} diagnostics. The first is the broad
all-integer row classification, containing \(108{,}642\) materials,
which provides a database-scale comparison with the original MP
output. The second is a matched composition baseline obtained by
enumerating the MP candidate oxidation states and requiring at least
one exact-neutral assignment with exactly one nonzero integer
oxidation state per element. This stricter, like-for-like baseline
contains \(89{,}374\) materials and is used for the mechanistic uplift
analysis in Table~\ref{tab:mixed-wyckoff}.

The comparison therefore separates four quantities: the broad MP
all-integer classification as a coverage benchmark, the matched MP
composition subset as the formula-level reference, atlas composition
mode as the like-for-like atlas comparison, and atlas Wyckoff mode as
the structural extension in which crystallographically distinct site
tokens are assigned independently.

\begin{figure}[t]
	\centering
	\includegraphics[width=0.92\columnwidth]{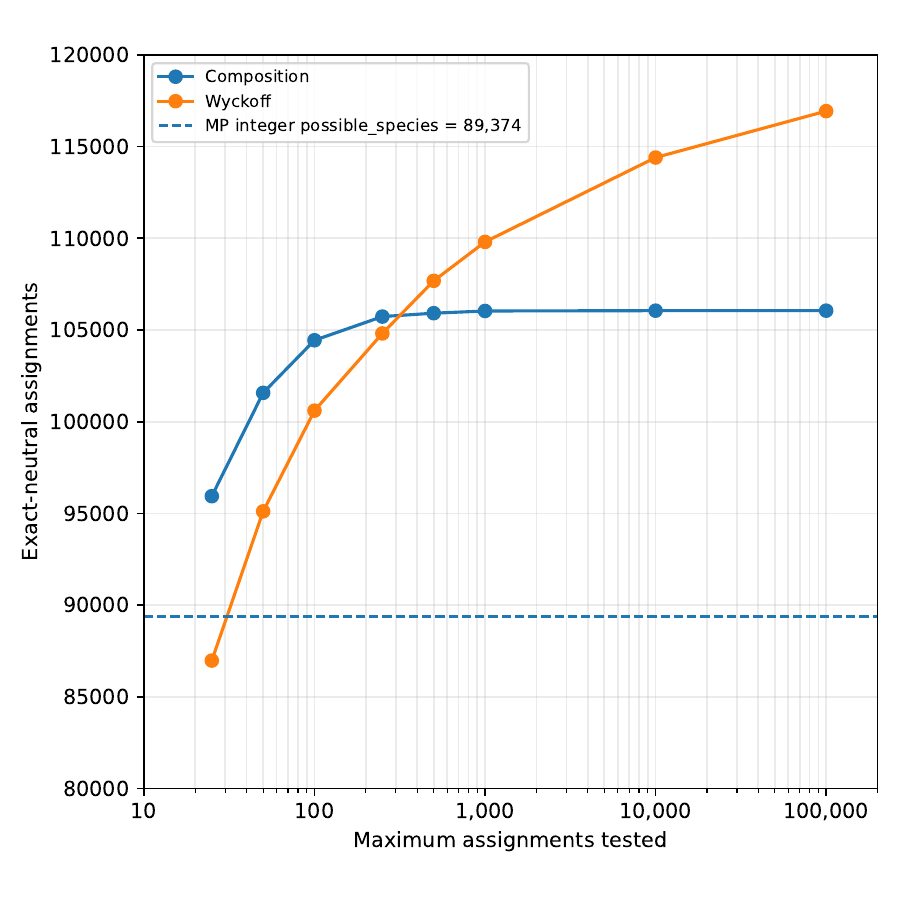}
	\caption{
		Atlas-prior oxidation-state assignment coverage compared with
		the broad MP all-integer \texttt{possible\_species} row classification. Composition mode assigns one oxidation state per element; Wyckoff mode assigns crystallographic site
		tokens independently using the same learned prior. At
		\(P_{\min}=0.01\) and
		\(N_{\mathrm{assign}}=10^{4}\), composition mode assigns
		\(106{,}053\) materials, close to the broad MP count of
		\(108{,}642\), while Wyckoff mode assigns \(114{,}403\),
		a \(5.3\%\) increase relative to that broad benchmark. The
		stricter matched MP composition baseline used for the
		mechanistic uplift analysis contains \(89{,}374\) materials
		and is reported in Table~\ref{tab:mixed-wyckoff}.
	}
	\label{fig:assignment_performance}
\end{figure}

Relative to the broad MP all-integer row classification,
composition-mode atlas coverage rises rapidly with the assignment cap
and reaches \(106{,}053\) materials, \(2.4\%\) below the broad MP
count of \(108{,}642\) (Fig.~\ref{fig:assignment_performance}).
Wyckoff mode reaches \(114{,}403\) materials, \(5.3\%\) above the
same broad benchmark at \(N_{\mathrm{assign}}=10^{4}\). Increasing
the cap to \(N_{\mathrm{assign}}=10^{5}\) yields only an additional
\(\sim 2.5\%\), indicating that most successful high-prior Wyckoff
assignments are recovered at the smaller budget.

The agreement between atlas composition coverage and the broad MP
count shows that the learned atlas provides a lightweight
representation of much of the formula-level oxidation-state
information encoded in the established MP workflow, despite using
only element-normalized probabilities and exact charge neutrality.
This is a database-scale coverage comparison rather than a
like-for-like comparison or a claim that the atlas supersedes the MP
assignments.

The matched MP subset provides the appropriate like-for-like
formula-level reference. Relative to its \(89{,}374\) materials,
atlas composition and Wyckoff modes increase assignment coverage by
\(18.7\%\) and \(28.0\%\), respectively. The composition comparison
isolates oxidation-state-library and candidate-enumeration effects. The additional Wyckoff comparison isolates the effect of
assigning crystallographically distinct site tokens independently.
The broad MP classification remains the reference for
Fig.~\ref{fig:assignment_performance}, while the matched subset is
used for the set-resolved and mechanistic analysis in
Table~\ref{tab:mixed-wyckoff}.

\subsection{Probability-threshold calibration and operating point}

The probability cutoff \(P_{\min}\) controls the retained oxidation-state tail (Table~\ref{tab:pmin_ox}). In composition mode, lowering the cutoff monotonically increases assignment coverage because the candidate space remains modest. In Wyckoff mode, coverage is maximal at \(P_{\min}=0.01\): admitting additional low-probability states increases the per-site candidate space and can prevent otherwise successful assignments from being reached within the fixed cap, \(N_{\mathrm{assign}}\). We therefore use \(P_{\min}=0.01\) and \(N_{\mathrm{assign}}=10^{4}\) as the operating point for the remaining atlas-application analyses.

The cutoff therefore acts as a practical regularization parameter.
Lowering \(P_{\min}\) admits additional low-probability states for
every independent Wyckoff/site variable, causing multiplicative
growth of the candidate space at fixed search budget. The optimum near \(P_{\min}=0.01\) retains most observed oxidation-state support while suppressing tails that otherwise reduce site-resolved search efficiency. Likewise, the modest improvement obtained by increasing the assignment cap to \(10^{5}\) indicates that the probability-ranked Cartesian traversal recovers most successful assignments within the smaller budget.

\begin{table*}[t]
	\caption{Oxidation state assignment success with decreasing oxidation state tail cutoff probability, for composition and Wyckoff grammar mode. $N_{\mathrm{assign}} = 10^{4}$. The total number of allowed oxidation states is $N_{\mathrm{ox}} = 336$.}
	\label{tab:pmin_ox}
	\begin{ruledtabular}
	\begin{tabular}{lccc}
		\(P_{\min}\) & ox states & Composition & Wyckoff \\
		\hline
		0.02  & 270 & 97,975  & 109,355 \\
		0.01  & 297 & 106,053 & 114,403 \\
		0.005 & 316 & 113,487 & 113,218 \\
		0.001 & 333 & 114,659 & 112,458 \\
	\end{tabular}
	\end{ruledtabular}
\end{table*}

\subsection{Wyckoff-uplift mechanism}

We next separated three questions: whether the atlas increases assignment coverage relative to the matched Materials Project (MP) baseline, whether the composition-level increase depends on the oxidation-state library, and whether the additional Wyckoff-mode assignments arise from site-resolved formal multivalence. A material was classified as having mixed-Wyckoff oxidation if at least one element occupied two or more Wyckoff/site tokens and received two or more distinct formal oxidation states in the Wyckoff-mode assignment. This diagnostic identifies formal site-resolved oxidation-state splitting, not direct evidence of experimentally localized electronic mixed valence.

Wyckoff resolution therefore adds a structurally meaningful degree of freedom rather than merely enlarging the search space: composition mode enforces one oxidation state per element, whereas Wyckoff mode can represent distinct formal oxidation states on crystallographically distinct sites of the same element.

\begin{table*}[t]
	\centering
	\small
	\setlength{\tabcolsep}{5pt}
	\caption{Assignment coverage, pairwise gains and losses, and the mixed-Wyckoff mechanism. The MP baseline includes materials for which the MP candidate species contain at least one charge-neutral assignment with exactly one nonzero integer oxidation state per element. Mixed-Wyckoff fractions are reported only for sets defined by successful Wyckoff-mode assignments, for which the diagnostic is evaluable for every material.}
	\label{tab:mixed-wyckoff}
	\begin{ruledtabular}
	\begin{tabular}{@{}lrrr@{}}
		Set & Materials & Mixed materials & Fraction \\
		\hline
		\multicolumn{4}{@{}l}{\textit{Overall assignment coverage}} \\
		MP matched composition & 89,374 & -- & -- \\
		Atlas composition & 106,053 & -- & -- \\
		Atlas Wyckoff & 114,403 & 36,966 & 32.3\% \\
		\hline
		\multicolumn{4}{@{}l}{\textit{Pairwise gains and losses}} \\
		Composition not MP & 19,525 & -- & -- \\
		MP not composition & 2,846 & -- & -- \\
		Wyckoff not composition & 14,665 & 14,662 & 99.98\% \\
		Composition not Wyckoff & 6,315 & -- & -- \\
		\hline
		\multicolumn{4}{@{}l}{\textit{Combined Wyckoff comparison with MP}} \\
		Wyckoff not MP & 31,514 & 21,900 & 69.5\% \\
		MP not Wyckoff & 6,485 & -- & -- \\
	\end{tabular}
	\end{ruledtabular}
\end{table*}

The first comparison establishes a general atlas uplift. Under the matched composition-level criterion, MP provides assignments for 89{,}374 materials, compared with 106,053 successes in atlas composition mode. The two sets are not nested: the atlas composition mode assigns 19,525 materials not recovered by MP, while MP assigns 2,846 materials not recovered by the atlas composition mode. The resulting net gain is therefore
\[
19{,}525 - 2{,}846 = 16{,}679,
\]
which equals the difference between the total success counts,
\[
106{,}053 - 89{,}374 = 16{,}679.
\]

Because this uplift is already present in composition mode, it cannot arise from Wyckoff-site resolution. Instead, it demonstrates a strong dependence on the oxidation-state candidate library and the associated enumeration procedure. The atlas uses the Greenwood--Earnshaw oxidation-state compilation, whereas the MP baseline is restricted to the oxidation states returned by the MP candidate-species heuristic. Among the 19,525 atlas-composition successes absent from the matched MP set, 9,763 correspond to missing or unparseable MP assignments, 7,346 have no charge-neutral MP solution using one oxidation state per element, 2,222 contain fractional MP oxidation states, and 194 require zero valence within the MP candidate set. Thus, the composition-level uplift primarily reflects broader or different oxidation-state coverage rather than structural resolution.

The second comparison isolates the effect of Wyckoff-site resolution. Wyckoff mode succeeds for 114,403 materials, compared with 106,053 for composition mode. Wyckoff mode recovers 14,665 materials not assigned in composition mode, while composition mode succeeds for 6,315 materials for which the finite Wyckoff search does not return a neutral assignment. The net Wyckoff gain is
\[
14{,}665 - 6{,}315 = 8{,}350,
\]
which equals
\[
114{,}403 - 106{,}053 = 8{,}350.
\]

The apparent non-nesting of the observed success sets is entirely attributable to finite search truncation. Composition mode is formally embedded in Wyckoff mode because a composition assignment is recovered by assigning the same oxidation state to every Wyckoff/site token of a given element. All 6,315 composition-success/Wyckoff-failure cases reached the search ceiling of \(N_{\mathrm{assign}}=10^{4}\), and the estimated Wyckoff candidate space exceeded this limit in every case. For these materials, the median estimated candidate-space size increased from 180 in composition mode to approximately \(3.74\times10^{12}\) in Wyckoff mode. Calculated separately for each material, the median Wyckoff-to-composition candidate-space expansion factor was approximately \(2.62\times10^{10}\). This value differs from the ratio of the two median candidate-space sizes because the median of per-material ratios is not generally equal to the ratio of medians.

Candidate tuples were evaluated in the direct Cartesian-product order of the probability-ranked oxidation-state lists for the individual Wyckoff/site variables. The search did not seed or otherwise prioritize composition-equivalent site-uniform assignments. Consequently, although the composition-level solution is contained in the full Wyckoff assignment space, it need not occur within the first \(N_{\mathrm{assign}}=10^{4}\) candidate tuples when the site-resolved space is combinatorially large. The composition-only cases therefore reflect combinatorial search truncation rather than reduced expressive capacity or an inconsistency in the Wyckoff representation.

The mechanism of the specifically structural uplift is nearly unambiguous. In the set
\[
S_{\mathrm{Wyckoff}} \setminus S_{\mathrm{comp}},
\]
14,662 of 14,665 materials, or \(99.98\%\), exhibit mixed-Wyckoff oxidation. Thus, nearly every material recovered only after introducing Wyckoff-site resolution contains at least one element assigned two or more distinct formal oxidation states on crystallographically distinct site tokens. Across all Wyckoff successes, 36,966 of 114,403 materials, or \(32.3\%\), exhibit mixed-Wyckoff oxidation. The increase from \(32.3\%\) overall to \(99.98\%\) in the Wyckoff-only set identifies formal site-resolved multivalence as the dominant mechanism of the Wyckoff-specific uplift. The near-universal mixed-Wyckoff character of the Wyckoff-only set shows that the structural uplift is not a generic counting artifact. It arises because site-resolved formal oxidation-state splitting can represent chemically plausible assignments that are inaccessible to a one-state-per-element composition model.

The combined Wyckoff-over-MP set contains 31,514 materials, of which 21,900, or \(69.5\%\), show mixed-Wyckoff oxidation. This set contains both contributions: materials recovered through the Greenwood--Earnshaw oxidation-state library already at the composition level and materials recovered only after site resolution. It therefore quantifies the total atlas advantage relative to MP, but it does not isolate a single mechanism.

Elements frequently involved among the 20 most common mixed-Wyckoff elements include the multivalent transition metals Mn, Fe, V, Co, Ni, Cu, Ti, Cr, Mo, and Nb, together with the main-group elements P, Ge, S, C, Sn, Si, B, N, Se, and Sb. Additional classically multivalent elements, including lanthanides such as Ce and Eu and heavier p-block elements such as Pb, Bi, Te, As, and Tl, also occur but at lower frequency. The transition-metal cases are consistent with familiar multivalent solid-state chemistry, while the main-group cases may also reflect chemically distinct local bonding environments, including cationic, anionic, and network-forming motifs. These assignments should therefore be interpreted as formal oxidation-state splitting rather than direct evidence of experimentally localized electronic mixed valence. Nevertheless, they occur in chemically plausible classes for which a single composition-level oxidation label can be too restrictive.

\subsection{Anion-conditioned atlases}

The O-, N-, and S-conditioned atlases were generated using the same staged procedure as the global atlas and are reported in the
Supplemental Material~\cite{SupplementalMaterial}. Oxygen-containing materials are dominated by \(\mathrm{O}^{-2}\). Nitrogen- and sulfur-containing subsets retain broader anion-state support because these elements occur in multiple bonding motifs. The Supplemental Material also reports the corresponding band-gap- and energy-above-hull-resolved sequential
level-0\(\rightarrow\)level-1 accounting tables using the same bin
definitions and outcome categories as Tables~\ref{tab:bandgap_accounting} and
\ref{tab:ehull_accounting}. These conditioned atlases are not used to optimize the main assignment comparison; they serve as chemically resolved diagnostics of how the global prior averages over distinct redox regimes.

\section{Conclusions}

We introduced a Wyckoff-resolved oxidation-state atlas constructed by staged level-0 and level-1 exact-neutral enumeration of Materials Project-derived structures. Applied as an empirical prior, the atlas assigns \(106{,}053\) materials in composition mode and \(114{,}403\) in Wyckoff mode. Composition-mode coverage approaches the broad MP all-integer benchmark, while comparison with the matched \(89{,}374\)-material MP baseline separates oxidation-state-library effects from the additional structural information introduced by
Wyckoff-site resolution.

Of the \(14{,}665\) materials recovered in Wyckoff mode but not in
composition mode, \(99.98\%\) exhibit distinct formal oxidation states for the same element on different site tokens. This identifies site-resolved formal multivalence as the dominant mechanism of the Wyckoff-specific uplift. The transparent CSV/Python workflow and standalone assignment utility therefore provide both a statistical summary of oxidation-state occurrence and a practical prior for chemically constrained structure decoration and materials discovery.

\begin{acknowledgments}
B.K. acknowledges support from the U.S. National Science Foundation under Award No.~2423992. The author also acknowledges the Materials Project for providing access to the computed materials data used as the starting point for this study.
\end{acknowledgments}

\section*{Conflict of Interest}
The author declares no conflict of interest.

\section*{Data Availability}
The serialized Wyckoff-grammar CSV derived from Materials Project structures, the global and anion-conditioned oxidation-state atlases, the standalone atlas-prior assignment utility, and the Python scripts used to generate the reported results are available at \url{https://github.com/boriskiefer/oxidation-state-atlas} under an MIT license.

The repository contains the scripts used to perform the staged level-0 and level-1 exact-neutral enumeration, accumulate the element-normalized oxidation-state probabilities, generate the figures and tables, apply the atlas-prior assignment utility in composition and Wyckoff modes, compare the atlas assignments with Materials Project \texttt{possible\_species} annotations, and analyze
the mixed-Wyckoff uplift mechanism. Example configuration files and command-line instructions are included.

The repository provides the serialized Wyckoff-grammar CSV used as the processed starting point for the reported analysis, together with the derived oxidation-state atlases and the scripts required to reproduce the reported results. The original Materials Project structures and underlying database records are not redistributed. They remain available from the Materials Project (\url{https://materialsproject.org}) subject to its terms of use.


\clearpage

\begin{widetext}
	\begin{center}
		{\large\bfseries Supplemental Material for}\\[0.5em]
		{\large\bfseries
			``Wyckoff-Resolved Oxidation-State Atlas and Anion-Conditioned Priors for Materials Discovery''
		}\\[1em]
		
		Boris Kiefer\\
		\textit{Department of Physics, New Mexico State University, Las Cruces, New Mexico 88003, USA}
	\end{center}
\end{widetext}

\renewcommand{\thesection}{S\arabic{section}}
\renewcommand{\thesubsection}{\thesection.\arabic{subsection}}
\renewcommand{\thefigure}{S\arabic{figure}}
\renewcommand{\theHfigure}{S\arabic{figure}}
\renewcommand{\thetable}{S\arabic{table}}
\renewcommand{\theHtable}{S\arabic{table}}
\renewcommand{\theequation}{S\arabic{equation}}

\setcounter{section}{0}
\setcounter{figure}{0}
\setcounter{table}{0}
\setcounter{equation}{0}

\section{Greenwood-Earnshaw oxidation-state table}\label{secA1}

The oxidation-state atlas was initialized from tabulated common and known oxidation states in Greenwood and Earnshaw's \emph{Chemistry of the Elements} \cite{GreenwoodEarnshaw1997}. Table~\ref{tab:ge_ox_atlas} lists the candidate states used in the staged enumeration workflow. Level-0 contains common nonzero oxidation states and defines the first-pass candidate set. Level-1 contains additional known nonzero oxidation states and is used only for materials not resolved at level-0. The zero oxidation state was excluded from both levels to emphasize formal ionic and redox-active oxidation chemistry and to avoid trivial charge-neutral assignments in metallic or strongly covalent systems.

\begin{center}
	\scriptsize
	\setlength{\tabcolsep}{0.25pt}
	\renewcommand{\arraystretch}{0.83}
	\setlength{\LTcapwidth}{0.95\linewidth}

    \begin{longtable}{@{}r @{\hspace{2pt}} lll@{}}
		\caption{Oxidation-state table used in this work. Level-0 lists \(N_{\mathrm{level}\,0}=157\) common oxidation states, and level-1 lists an additional \(N_{\mathrm{level}\,1}=212\) known oxidation states, initialized from Greenwood and Earnshaw \cite{GreenwoodEarnshaw1997}.}
		\label{tab:ge_ox_atlas}\\
		
		\hline \hline
		\(Z\) & Element & level-0 (common) & level-1 (known) \\
		\hline
		\endfirsthead
		
		\caption[]{Oxidation-state table used in this work, continued.}\\
		\hline \hline
		\(Z\) & Element & level-0 (common) & level-1 (known) \\
		\hline
		\endhead
		
		\hline
		\endfoot
		
		\hline \hline
		\endlastfoot
		
		1 & H & $-1, +1$ &  \\
		2 & He &  &  \\
		3 & Li & $+1$ & $-1$ \\
		4 & Be & $+2$ &  \\
		5 & B & $+3$ & $+1, +2$ \\
		6 & C & $-4, +4$ & $-3, -2, -1, +1, +2, +3$ \\
		7 & N & $-3, +3, +5$ & $-2, -1, +1, +2, +4$ \\
		8 & O & $-2$ & $-1, +1, +2$ \\
		9 & F & $-1$ &  \\
		10 & Ne &  &  \\
		11 & Na & $+1$ & $-1$ \\
		12 & Mg & $+2$ &  \\
		13 & Al & $+3$ & $+1$ \\
		14 & Si & $-4, +4$ & $+1, +2, +3, -3, -2, -1$ \\
		15 & P & $-3, +3, +5$ & $+1, +2, +4, -2, -1$ \\
		16 & S & $-2, +2, +4, +6$ & $-1, +1, +3, +5$ \\
		17 & Cl & $-1, +1, +3, +5, +7$ & $+2, +4, +6$ \\
		18 & Ar &  &  \\
		19 & K & $+1$ &  \\
		20 & Ca & $+2$ &  \\
		21 & Sc & $+3$ & $+1, +2$ \\
		22 & Ti & $+4$ & $+2, +3, -1$ \\
		23 & V & $+5$ & $+1, +2, +3, +4, -1$ \\
		24 & Cr & $+3, +6$ & $+1, +2, +4, +5, -2, -1$ \\
		25 & Mn & $+2, +4, +7$ & $+1, +3, +5, +6, -3, -2, -1$ \\
		26 & Fe & $+2, +3$ & $+1, +4, +5, +6, +7, -2, -1$ \\
		27 & Co & $+2, +3$ & $+1, +4, +5, -1$ \\
		28 & Ni & $+2$ & $+1, +3, +4, -1$ \\
		29 & Cu & $+2$ & $+1, +3, +4$ \\
		30 & Zn & $+2$ &  \\
		31 & Ga & $+3$ & $+1, +2$ \\
		32 & Ge & $+2, +4, -4$ & $+1, +3$ \\
		33 & As & $+3, +5, -3$ & $+2$ \\
		34 & Se & $-2, +2, +4, +6$ & $-1, +1, +3, +5$ \\
		35 & Br & $-1, +1, +3, +5$ & $+4$ \\
		36 & Kr & $+2$ &  \\
		37 & Rb & $+1$ &  \\
		38 & Sr & $+2$ &  \\
		39 & Y & $+3$ & $+2$ \\
		40 & Zr & $+4$ & $+1, +2, +3$ \\
		41 & Nb & $+5$ & $-1, +2, +3, +4$ \\
		42 & Mo & $+4, +6$ & $-2, -1, +1, +2, +3, +5$ \\
		43 & Tc & $+4, +7$ & $-3, -1, +1, +2, +3, +5, +6$ \\
		44 & Ru & $+3, +4$ & $-2, +1, +2, +5, +6, +7, +8$ \\
		45 & Rh & $+3$ & $-1, +1, +2, +4, +5, +6$ \\
		46 & Pd & $+2, +4$ &  \\
		47 & Ag & $+1$ & $+2, +3$ \\
		48 & Cd & $+2$ &  \\
		49 & In & $+3$ & $+1, +2$ \\
		50 & Sn & $-4, +2, +4$ &  \\
		51 & Sb & $-3, +3, +5$ &  \\
		52 & Te & $-2, +2, +4, +6$ & $+5$ \\
		53 & I & $-1, +1, +3, +5, +7$ &  \\
		54 & Xe & $+2, +4, +6$ & $+8$ \\
		55 & Cs & $+1$ &  \\
		56 & Ba & $+2$ &  \\
		57 & La & $+3$ & $+2$ \\
		58 & Ce & $+3, +4$ & $+2$ \\
		59 & Pr & $+3$ & $+2, +4$ \\
		60 & Nd & $+3$ & $+2$ \\
		61 & Pm & $+3$ &  \\
		62 & Sm & $+3$ & $+2$ \\
		63 & Eu & $+2, +3$ &  \\
		64 & Gd & $+3$ & $+1, +2$ \\
		65 & Tb & $+3$ & $+1, +4$ \\
		66 & Dy & $+3$ & $+2$ \\
		67 & Ho & $+3$ &  \\
		68 & Er & $+3$ &  \\
		69 & Tm & $+3$ & $+2$ \\
		70 & Yb & $+3$ & $+2$ \\
		71 & Lu & $+3$ &  \\
		72 & Hf & $+4$ & $+2, +3$ \\
		73 & Ta & $+5$ & $+2, +3, +4$ \\
		74 & W & $+4, +6$ & $-2, -1, +1, +2, +3, +5$ \\
		75 & Re & $+4$ & $-3, -1, +1, +2, +3, +5, +6, +7$ \\
		76 & Os & $+4$ & $-2, +1, +2, +3, +5, +6, +7, +8$ \\
		77 & Ir & $+3, +4$ & $-1, +1, +2, +5, +6$ \\
		78 & Pt & $+2, +4$ & $+5, +6$ \\
		79 & Au & $+3$ & $-1, +1, +2$ \\
		80 & Hg & $+1, +2$ &  \\
		81 & Tl & $+1, +3$ &  \\
		82 & Pb & $+2, +4$ & $-4$ \\
		83 & Bi & $+3$ & $-3, +5$ \\
		84 & Po & $-2, +2, +4$ & $+6$ \\
		85 & At & $-1, +1$ & $+3, +5, +7$ \\
		86 & Rn &  &  \\
		87 & Fr & $+1$ &  \\
		88 & Ra & $+2$ &  \\
		89 & Ac & $+3$ &  \\
		90 & Th & $+4$ & $+2, +3$ \\
		91 & Pa & $+5$ & $+3, +4$ \\
		92 & U & $+6$ & $+3, +4, +5$ \\
		93 & Np & $+5$ & $+3, +4, +6, +7$ \\
		94 & Pu & $+4$ & $+3, +5, +6, +7$ \\
	\end{longtable}
\end{center}

\section{Anion-conditioned atlases}

The O-, N-, and S-conditioned atlases were generated by filtering the Materials Project-derived structure set to entries containing the specified anion and then applying the same staged level-0\(\rightarrow\)level-1 exact-neutral enumeration and Wyckoff-site tallying procedure used for the all-compound atlas. These heatmaps provide a chemical-family-resolved view of the global oxidation-state prior. They use the same element-normalized probability scale, level-0/level-1 marker convention, and Wyckoff-site mass normalization as the main-text ALL atlas.

\clearpage

\begin{center}
\begin{minipage}{0.95\textwidth}
	\centering
	\includegraphics[width=\textwidth,height=0.85\textheight]{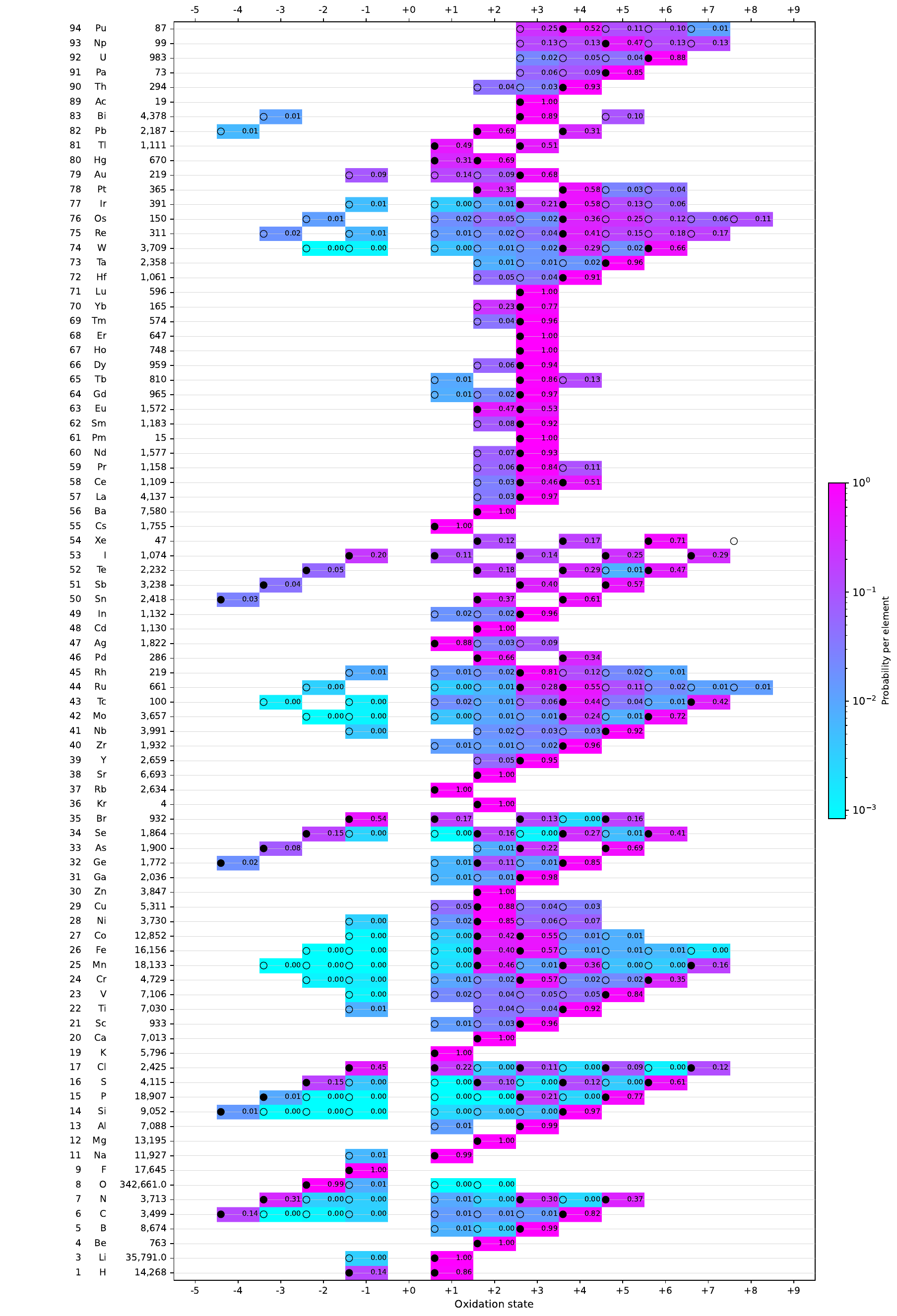}
	
 \vspace{0.5em}

\refstepcounter{figure}
\raggedright

		\noindent
		\textbf{FIG. \thefigure.}
		Oxygen-conditioned oxidation-state atlas constructed from the subset of Materials Project-derived structures containing oxygen.
		Each row corresponds to an element and each column to an integer oxidation state.
		Colors denote the element-normalized probability \(P(q \mid Z)\) obtained from the staged level-0\(\rightarrow\)level-1 exact-neutral atlas construction, using Wyckoff-site mass as the default weighting.
		Filled circles mark level-0 oxidation states, and open circles mark level-1 oxidation states.
		The left-hand annotation reports the total Wyckoff-site occurrence for each element within the oxygen-conditioned subset.
		\label{fig:heatmap_O}
	\end{minipage}
\end{center}

\clearpage

\begin{center}
	\begin{minipage}{0.95\textwidth}
	\includegraphics[width=\textwidth,height=0.85\textheight]{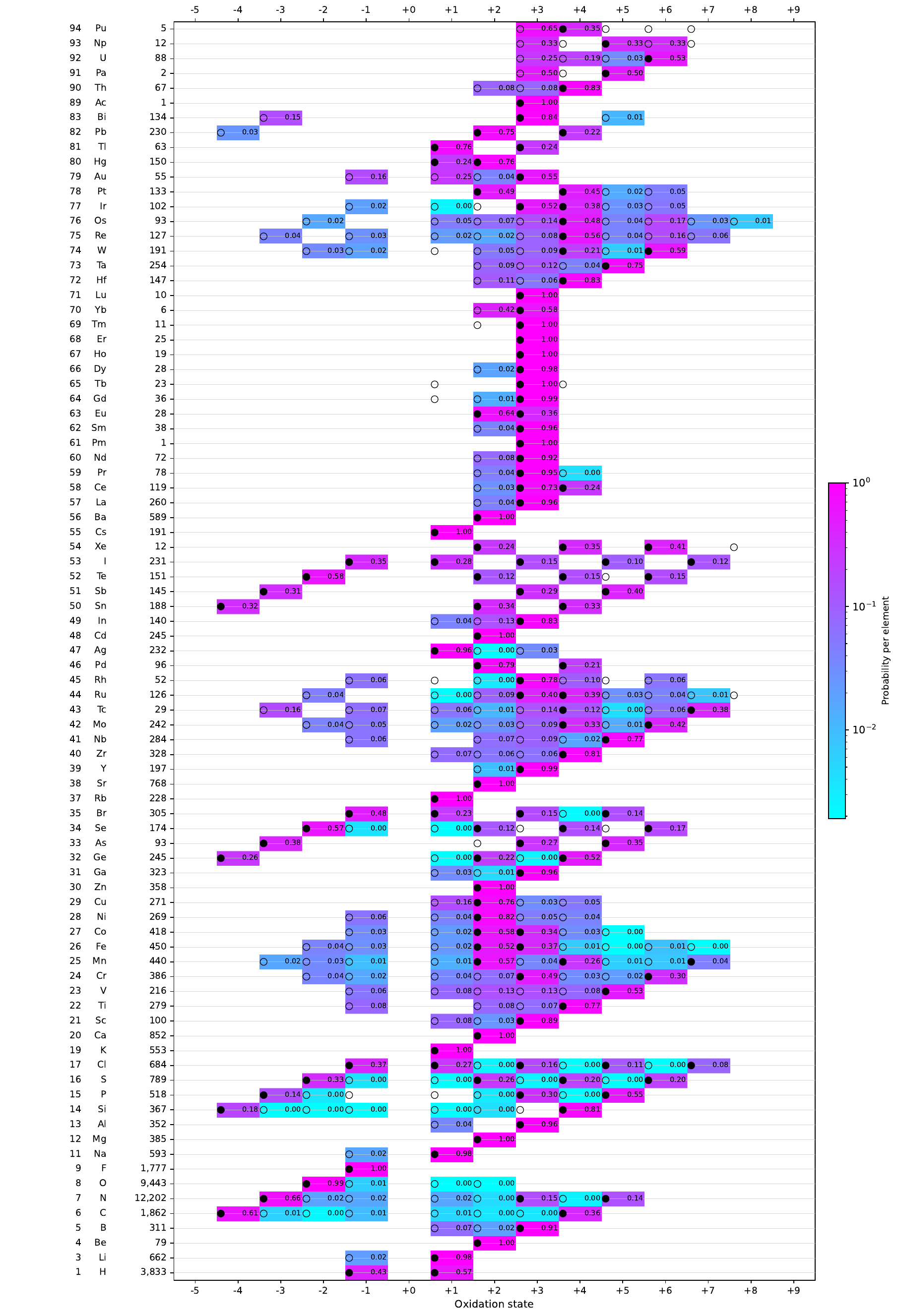}

 \vspace{0.5em}

\refstepcounter{figure}
\raggedright

		\noindent
		\textbf{FIG. \thefigure.}
		Nitrogen-conditioned oxidation-state atlas constructed from the subset of Materials Project-derived structures containing nitrogen. Each row corresponds to an element and each column to an integer oxidation state.
		Colors denote the element-normalized probability \(P(q \mid Z)\) obtained from the staged level-0\(\rightarrow\)level-1 exact-neutral atlas construction, using Wyckoff-site mass as the default weighting.
		Filled circles mark level-0 oxidation states, and open circles mark level-1 oxidation states.
		The left-hand annotation reports the total Wyckoff-site occurrence for each element within the nitrogen-conditioned subset.
		\label{fig:heatmap_N}		
	\end{minipage}
\end{center}

\clearpage

\begin{center}
	\begin{minipage}{0.95\textwidth}
	\includegraphics[width=\textwidth,height=0.85\textheight]{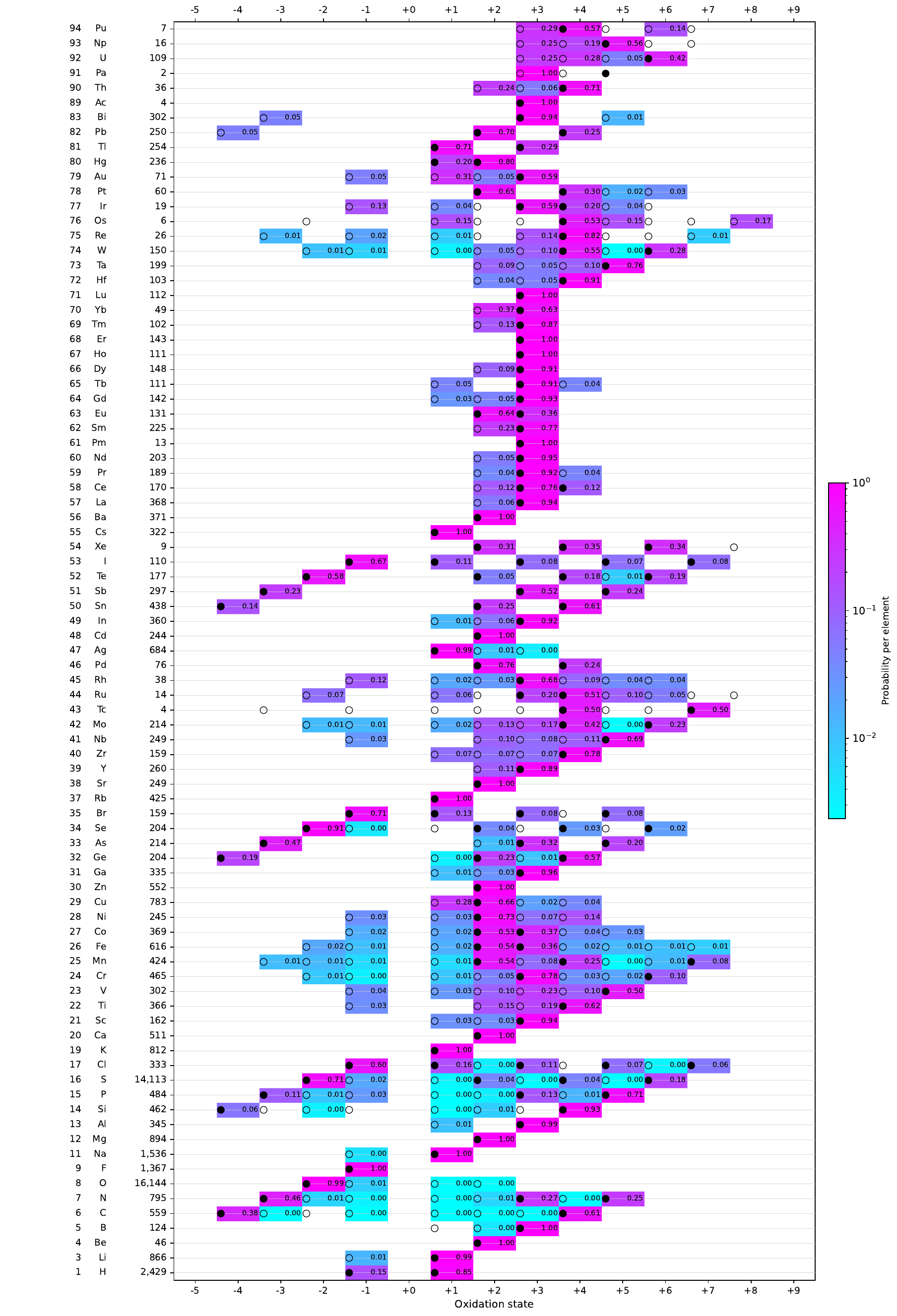}
	
 \vspace{0.5em}

\refstepcounter{figure}
\raggedright

		\noindent
		\textbf{FIG. \thefigure.}
		Sulfur-conditioned oxidation-state atlas constructed from the subset of Materials Project-derived structures containing sulfur.
		Each row corresponds to an element and each column to an integer oxidation state. Colors denote the element-normalized probability \(P(q \mid Z)\) obtained from the staged level-0\(\rightarrow\)level-1 exact-neutral atlas construction, using Wyckoff-site mass as the default weighting.
		Filled circles mark level-0 oxidation states, and open circles mark level-1 oxidation states.
		The left-hand annotation reports the total Wyckoff-site occurrence for each element within the sulfur-conditioned subset.
		\label{fig:heatmap_S}		
	\end{minipage}
\end{center}

\clearpage

\section{Anion-resolved property accounting}

Tables~\ref{tab:SI_O_bandgap_accounting}--\ref{tab:SI_S_ehull_accounting} report anion-resolved sequential level-0\(\rightarrow\)level-1 construction accounting for the O-, N-, and S-conditioned atlases. The same replacement logic, property bins, and outcome categories are used as in the all-compound main-text tables: level-0 successes are retained, while level-0 failures are replaced by their level-1 outcomes. Percentages are normalized within each property bin. These tables provide chemical-family-resolved diagnostics of atlas coverage and dominant unresolved categories.

\begin{widetext}
\begin{center}
	\refstepcounter{table}
	\begin{minipage}{0.95\textwidth}
		\textbf{TABLE \thetable.}
		Band-gap-resolved accounting of oxidation-state enumeration outcomes for the \textbf{oxygen}-conditioned sequential level-0\(\rightarrow\)level-1 atlas at \(N_{\mathrm{assign}}=10{,}000\), using the assignment-cap definition described in the Methods section of the main text. Level-0 is applied first, and level-1 is applied only to entries that fail to obtain an exact-neutral level-0 assignment; therefore, level-0 successes are retained and level-0 failures are replaced by level-1 outcomes. Percentages are normalized within each band-gap bin and sum to \(100\%\). The ``Other'' category groups minor bookkeeping outcomes.
		\label{tab:SI_O_bandgap_accounting}
		
		\vspace{0.5em}

 	\centering
	\begin{ruledtabular}
		\begin{tabular}{lrrrrr}
			Band-gap bin & $N$ & Success & No solution & Limit & Other \\
			\hline
			$E_g \le 10^{-5}$        & 23,157 & 61.2 & 3.3 & 35.1 & 0.4 \\
			$10^{-5} < E_g \le 1$    & 18,253 & 55.4 & 0.8 & 43.5 & 0.3 \\
			$1 < E_g \le 2$          & 12,809 & 67.1 & 0.1 & 32.7 & 0.1 \\
			$E_g > 2$                & 27,966 & 78.3 & 0.0 & 21.7 & 0.0 \\
		\end{tabular}
	\end{ruledtabular}

\end{minipage}
\end{center}
\end{widetext}

\begin{widetext}
	\begin{center}
		\refstepcounter{table}
		\begin{minipage}{0.95\textwidth}
		\textbf{TABLE \thetable.}
		Energy-above-hull-resolved accounting of oxidation-state enumeration outcomes for the \textbf{oxygen}-conditioned sequential level-0\(\rightarrow\)level-1 atlas at \(N_{\mathrm{assign}}=10{,}000\), using the assignment-cap definition described in the Methods section of the main text. Level-0 is applied first, and level-1 is applied only to entries that fail to obtain an exact-neutral level-0 assignment; therefore, level-0 successes are retained and level-0 failures are replaced by level-1 outcomes. Percentages are normalized within each \(E_{\mathrm{hull}}\) bin and sum to \(100\%\). The ``Other'' category groups minor bookkeeping outcomes.
		\label{tab:SI_O_ehull_accounting}
		
		\vspace{0.5em}
		
		\centering
	\begin{ruledtabular}
		\begin{tabular}{lrrrrr}
			$E_{\mathrm{hull}}$ bin & $N$ & Success & No solution & Limit & Other \\
			\hline
			$E_{\mathrm{hull}} \le 0.05$        & 34,969 & 75.7 & 0.7 & 23.5 & 0.1 \\
			$0.05 < E_{\mathrm{hull}} \le 0.10$ & 17,679 & 61.0 & 0.7 & 38.2 & 0.1 \\
			$0.10 < E_{\mathrm{hull}} \le 0.25$ & 17,134 & 60.4 & 1.2 & 38.2 & 0.1 \\
			$E_{\mathrm{hull}} > 0.25$          & 12,403 & 57.7 & 2.7 & 38.7 & 0.9 \\
		\end{tabular}
	\end{ruledtabular}
\end{minipage}
\end{center}
\end{widetext}

\begin{widetext}
	\begin{center}
		\refstepcounter{table}
		\begin{minipage}{0.95\textwidth}
		\textbf{TABLE \thetable.}
		Band-gap-resolved accounting of oxidation-state enumeration outcomes for the \textbf{nitrogen}-conditioned sequential level-0\(\rightarrow\)level-1 atlas at \(N_{\mathrm{assign}}=10{,}000\), using the assignment-cap definition described in the Methods section of the main text. Level-0 is applied first, and level-1 is applied only to entries that fail to obtain an exact-neutral level-0 assignment; therefore, level-0 successes are retained and level-0 failures are replaced by level-1 outcomes. Percentages are normalized within each band-gap bin and sum to \(100\%\). The ``Other'' category groups minor bookkeeping outcomes.
		\label{tab:SI_N_bandgap_accounting}
		
		\vspace{0.5em}
		
		\centering
		\begin{ruledtabular}
		\begin{tabular}{lrrrrr}
			Band-gap bin & $N$ & Success & No solution & Limit & Other \\
			\hline
			$E_g \le 10^{-5}$        & 4,117 & 68.7 & 3.4 & 27.4 & 0.5 \\
			$10^{-5} < E_g \le 1$    & 2,169 & 51.8 & 0.3 & 47.5 & 0.3 \\
			$1 < E_g \le 2$          & 1,472 & 59.4 & 0.1 & 40.4 & 0.1 \\
			$E_g > 2$                & 3,679 & 51.2 & 0.0 & 48.2 & 0.6 \\
		\end{tabular}
		\end{ruledtabular}
\end{minipage}
\end{center}
\end{widetext}

\begin{widetext}
	\begin{center}
		\refstepcounter{table}
		\begin{minipage}{0.95\textwidth}
		\textbf{TABLE \thetable.}
		Energy-above-hull-resolved accounting of oxidation-state enumeration outcomes for the \textbf{nitrogen}-conditioned sequential level-0\(\rightarrow\)level-1 atlas at \(N_{\mathrm{assign}}=10{,}000\), using the assignment-cap definition described in the Methods section of the main text. Level-0 is applied first, and level-1 is applied only to entries that fail to obtain an exact-neutral level-0 assignment; therefore, level-0 successes are retained and level-0 failures are replaced by level-1 outcomes. Percentages are normalized within each \(E_{\mathrm{hull}}\) bin and sum to \(100\%\). The ``Other'' category groups minor bookkeeping outcomes.
		\label{tab:SI_N_ehull_accounting}
		
		\vspace{0.5em}
		
		\centering
		\begin{ruledtabular}
		\begin{tabular}{lrrrrr}
			$E_{\mathrm{hull}}$ bin & $N$ & Success & No solution & Limit & Other \\
			\hline
			$E_{\mathrm{hull}} \le 0.05$        & 3,140 & 65.9 & 2.8 & 30.5 & 0.7 \\
			$0.05 < E_{\mathrm{hull}} \le 0.10$ & 1,235 & 55.6 & 0.2 & 44.1 & 0.1 \\
			$0.10 < E_{\mathrm{hull}} \le 0.25$ & 2,554 & 53.5 & 0.5 & 46.0 & 0.0 \\
			$E_{\mathrm{hull}} > 0.25$          & 4,508 & 57.5 & 1.0 & 41.0 & 0.6 \\
		\end{tabular}
		\end{ruledtabular}
\end{minipage}
\end{center}
\end{widetext}

\begin{widetext}
	\begin{center}
		\refstepcounter{table}
		\begin{minipage}{0.95\textwidth}
		\textbf{TABLE \thetable.}
		Band-gap-resolved accounting of oxidation-state enumeration outcomes for the \textbf{sulfur}-conditioned sequential level-0\(\rightarrow\)level-1 atlas at \(N_{\mathrm{assign}}=10{,}000\), using the assignment-cap definition described in the Methods section of the main text. Level-0 is applied first, and level-1 is applied only to entries that fail to obtain an exact-neutral level-0 assignment; therefore, level-0 successes are retained and level-0 failures are replaced by level-1 outcomes. Percentages are normalized within each band-gap bin and sum to \(100\%\). The ``Other'' category groups minor bookkeeping outcomes.
		\label{tab:SI_S_bandgap_accounting}
		
		\vspace{0.5em}
		
		\centering
		
		\begin{ruledtabular}
		\begin{tabular}{lrrrrr}
			Band-gap bin & $N$ & Success & No solution & Limit & Other \\
			\hline
			$E_g \le 10^{-5}$        & 4,503 & 50.0 & 5.4 & 44.2 & 0.4 \\
			$10^{-5} < E_g \le 1$    & 2,902 & 45.2 & 0.6 & 54.1 & 0.2 \\
			$1 < E_g \le 2$          & 2,909 & 46.1 & 0.1 & 53.6 & 0.3 \\
			$E_g > 2$                & 4,996 & 46.2 & 0.0 & 53.3 & 0.5 \\
		\end{tabular}
		\end{ruledtabular}
\end{minipage}
\end{center}
\end{widetext}

\begin{widetext}
	\begin{center}
		\refstepcounter{table}
		\begin{minipage}{0.95\textwidth}
		\textbf{TABLE \thetable.}
		Energy-above-hull-resolved accounting of oxidation-state enumeration outcomes for the \textbf{sulfur}-conditioned sequential level-0\(\rightarrow\)level-1 atlas at \(N_{\mathrm{assign}}=10{,}000\), using the assignment-cap definition described in the Methods section of the main text. Level-0 is applied first, and level-1 is applied only to entries that fail to obtain an exact-neutral level-0 assignment; therefore, level-0 successes are retained and level-0 failures are replaced by level-1 outcomes. Percentages are normalized within each \(E_{\mathrm{hull}}\) bin and sum to \(100\%\). The ``Other'' category groups minor bookkeeping outcomes.
		\label{tab:SI_S_ehull_accounting}
		
		\vspace{0.5em}
		
		\centering
		
		\begin{ruledtabular}
		\begin{tabular}{lrrrrr}
			$E_{\mathrm{hull}}$ bin & $N$ & Success & No solution & Limit & Other \\
			\hline
			$E_{\mathrm{hull}} \le 0.05$        & 7,211 & 57.3 & 1.7 & 40.5 & 0.5 \\
			$0.05 < E_{\mathrm{hull}} \le 0.10$ & 2,755 & 35.1 & 1.3 & 63.4 & 0.1 \\
			$0.10 < E_{\mathrm{hull}} \le 0.25$ & 3,095 & 36.2 & 1.2 & 62.6 & 0.0 \\
			$E_{\mathrm{hull}} > 0.25$          & 2,249 & 44.0 & 3.0 & 52.2 & 0.8 \\
		\end{tabular}
		\end{ruledtabular}
\end{minipage}
\end{center}
\end{widetext}

\section{Application of the oxidation-state assignment utility}
The examples in Table~\ref{tab:SI_assignment_examples} illustrate four characteristic outcomes. For
\(\mathrm{MgO}\), composition mode returns a single dominant conventional assignment, \(\mathrm{Mg}^{+2}/\mathrm{O}^{-2}\). For \(\mathrm{SnO}_{2}\), the utility returns and ranks two exact-neutral solutions: the conventional \(\mathrm{Sn}^{+4}/\mathrm{O}^{-2}\) assignment and the lower-ranked formal \(\mathrm{Sn}^{+2}/\mathrm{O}^{-1}\) assignment. Composition mode does not resolve \(\mathrm{Fe}_{3}\mathrm{O}_{4}\) because it assigns only one oxidation state per element. In contrast, Wyckoff mode resolves \(\mathrm{Fe}_{3}\mathrm{O}_{4}\) by assigning different oxidation states to the two crystallographically distinct Fe site tokens. The normalized weights quantify the relative ranking only among the neutral solutions returned for each input under the default search settings. 

Representative calculations are executed as
\begin{verbatim}
	python step_7_use_atlas.py --mode composition \
	--input "MgO"
	
	python step_7_use_atlas.py --mode composition \
	--input "Fe3O4"
	
	python step_7_use_atlas.py --mode composition \
	--input "SnO2"
	
	python step_7_use_atlas.py --mode wyckoff \
	--input "Fe:8:a:Fe1;Fe:16:d:Fe2;O:32:e:O1"
\end{verbatim}

\begin{widetext}
	\begin{center}
		\refstepcounter{table}
		\begin{minipage}{0.95\textwidth}
		\textbf{TABLE \thetable.}
		Representative outputs of the oxidation-state assignment utility using the default ALL atlas and search settings. The normalized weight gives the relative weight among exact-neutral assignments returned for the same input, and \texttt{min\_site\_probability} is the smallest constituent atlas probability in the assignment.
		\label{tab:SI_assignment_examples}
		
		\vspace{0.5em}
		
		\centering
		\begin{ruledtabular}
		\begin{tabular}{llllrr}
			Input & Mode & Rank & Assignment & Normalized weight & min\_site\_probability \\
			\hline
			\(\mathrm{MgO}\) &
			composition & 1 &
			\(\mathrm{Mg}^{+2},\ \mathrm{O}^{-2}\) &
			1.000 & 0.991 \\
			\hline
			
			\(\mathrm{SnO}_{2}\) &
			composition & 1 &
			\(\mathrm{Sn}^{+4},\ \mathrm{O}^{-2}\) &
			0.995 & 0.409 \\
			
			\(\mathrm{SnO}_{2}\) &
			composition & 2 &
			\(\mathrm{Sn}^{+2},\ \mathrm{O}^{-1}\) &
			0.005 & 0.008 \\
			\hline
			
			\(\mathrm{Fe}_{3}\mathrm{O}_{4}\) &
			composition & -- & 
			no exact-neutral & -- & -- \\
			&  &  & assignment &  &  \\
			\hline
			
			\(\mathrm{Fe}_{3}\mathrm{O}_{4}\) &
			Wyckoff & 1 &
			\(\mathrm{Fe1}^{+2},\ \mathrm{Fe2}^{+3},
			\ \mathrm{O1}^{-2}\) &
			0.980 & 0.397 \\
			
			\(\mathrm{Fe}_{3}\mathrm{O}_{4}\) &
			Wyckoff & 2 &
			\(\mathrm{Fe1}^{+4},\ \mathrm{Fe2}^{+2}, \ \mathrm{O1}^{-2}\) &
			0.020 &
			0.011 \\
			
			\(\mathrm{Fe}_{3}\mathrm{O}_{4}\) &
			Wyckoff & 3 &
			\(\mathrm{Fe1}^{+6},\ \mathrm{Fe2}^{+1},\ \mathrm{O1}^{-2}\) &
			0.000 & 0.007 \\
			
			\(\mathrm{Fe}_{3}\mathrm{O}_{4}\) &
			Wyckoff & 4 &
			\(\mathrm{Fe1}^{+2},\ \mathrm{Fe2}^{+1},\
			\mathrm{O1}^{-1}\) &
			0.000 & 0.007
		\end{tabular}
		\end{ruledtabular}
\end{minipage}
\end{center}
\end{widetext}

\clearpage

\bibliography{oxatlas}

\end{document}